\documentclass[12pt, preprint]{aastex}
\usepackage[dvips]{color}

\def\ltsima{$\;\buildrel < \over \sim \;$}
\def\simlt{\lower.5ex \hbox{\ltsima}}
\def\gtsima{$\;\buildrel > \over \sim \;$}
\def\simgt{\lower.5ex \hbox{\gtsima}}

\shorttitle{Collisional excitation of far-IR CO emissions}

\shortauthors{Neufeld}

\begin{document}

\title{Collisional excitation of far-infrared line emissions from warm interstellar carbon monoxide (CO)}
\author{David A.~Neufeld}
\affil{Department of Physics and Astronomy, Johns Hopkins University,
3400~North~Charles~Street, Baltimore, MD 21218}

\begin{abstract}

Motivated by recent observations with {\it Herschel}/PACS, and the availability of new rate coefficients for the collisional excitation of CO (Yang et al.\ 2010), the excitation of warm astrophysical CO is revisited with the use of numerical and analytic methods.
For the case of an isothermal medium, results have been obtained for a wide range of gas temperatures (100 to 5000~K) and H$_2$ densities ($10^3 - 10^9 \, \rm cm^{-3}$), and presented in the form of rotational diagrams, in which the logarithm of the column density per magnetic substate, log~$(N_J/g_J)$, is plotted for each state, as a function of its energy, $E_J$.  For rotational transitions in the wavelength range accessible to {\it Herschel}/PACS, such diagrams are nearly linear when $n({\rm H}_2)\ge 10^8$~$\rm cm^{-3}$.  When $n({\rm H}_2) \sim 10^{6.8} - 10^8$~$\rm cm^{-3}$, they exhibit significant negative curvature, whereas when $n({\rm H}_2) \le 10^{4.8}$~$\rm cm^{-3}$, the curvature is uniformly positive throughout the PACS-accessible range.  
Thus, the observation of a positively-curved CO rotational diagram does not {\it necessarily} require the presence of multiple temperature components.   Indeed, for some sources observed with {\it Herschel}/PACS, the CO rotational diagrams show a modest positive curvature that can be explained by a single isothermal component.  Typically, the required physical parameters are densities in the $10^4 - 10^5$~cm$^{-3}$ range and temperatures close to the maximum at which CO can survive.  Other sources exhibit rotational diagrams with more curvature than can be accounted for by a single temperature component.  
For the case of a medium with a power-law distribution of gas temperatures, $dN/dT \propto T^{-b}$,  results have been obtained for H$_2$ densities $10^3 - 10^9 \, \rm cm^{-3}$ and power-law indices, $b$, in the range 1 to 5; such a medium can account for a CO rotational diagram that is more positively curved than any resulting from an isothermal medium.

\end{abstract}


\section{Introduction}

Over the past thirty years, far-infrared line emissions from warm carbon monoxide have been observed from a succession of airborne and satellite observatories: the Kuiper Airborne Observatory (e.g.\ Watson et al.\ 1985) the {\it Infrared Space Observatory} ({\it ISO}; e.g.\ Lerate et al.\ 2006), the {\it Herschel Space Observatory} (e.g. Herczeg et al.\ 2011) and the Stratospheric Observatory for Infrared Astronomy (SOFIA).  Such observations have provided an invaluable probe of warm molecular gas that has been heated by interstellar shock waves and in photodissociation regions (PDRs).   In the past three years, observations performed with {\it Herschel} -- and in particular its PACS spectrometer -- have yielded a large data set of far-infrared CO observations, providing detections of pure rotational transitions ranging from $J = 13 -12$ (at 200.272~$\mu$m) to $J=49 - 48$ (at 53.897~$\mu$m).  Measurements of the ladder of CO rotational transitions are conveniently represented by a rotational diagram, in which the logarithm of the column density per magnetic substate, ln~$(N_J/g_J)$, is plotted for each state, as a function of its energy, $E_J$.  This representation is motivated by the form of the Boltzmann factor, which leads -- for the case of an isothermal medium in local thermodynamic equilibrium (LTE) at temperature, $T$ -- to a straight line for which the slope is $-(1/kT)$; in this case, the y-intercept yields the total CO density divided by the partition function.  More generally, it is convenient to define a {\it rotational temperature},
$$T_{\rm rot} \equiv -(k\,{\rm dln}\,[N_J/g_J]/dE_J)^{-1}. \eqno(1)$$
which -- in the non-LTE case -- may show a variation with $E_J$.

Two features of the rotational diagrams obtained from {\it Herschel}/PACS observations are particularly noteworthy: (1) the plots typically exhibit positive curvature, with $T_{\rm rot}$
increasing monotonically with $E_J$; and (2) the value of $T_{\rm rot}$ determined from the {\it lowest} transitions accessible to {\it Herschel}/PACS is typically close to 300~K (e.g.\ van Kempen et al.\ 2010a,b; Fich et al.\ 2010; Herczeg et al.\ 2011).

The far-infrared emission spectra expected from warm CO have been modeled extensively in the literature (e.g.\ McKee et al.\ 1982, Viscuso \& Chernoff 1988).   The present study is motivated by recent observations with {\it Herschel}/PACS, as well as by the availability of new rate coefficients computed by Yang et al.\ (2011) for the collisional excitation of CO in collisions with H$_2$.  I will revisit the excitation of warm CO with the use of numerical and analytic solutions to the equations of statistical equilibrium for the level populations.  The calculational method is discussed in \S 2, including the techniques used to extrapolate the collisional rate coefficients to transitions lying higher than any considered by Yang et al.\ (2011).  In \S 3 and \S 4, the predicted rotational diagrams are presented for two cases: an isothermal medium, and a medium in which a {\it distribution} of gas temperatures is present.  Certain limiting behaviors are accounted for with an analytic treatment, which yields excellent agreement with the numerical results.  A brief summary follows in \S 5.

\section{Calculational method}

\subsection{Equations of statistical equilibrium}

Using methods identical to those described in Neufeld \& Melnick (1993), the equations of statistical equilibrium were solved to obtain the steady-state populations for 81 rotational states of CO (i.e.\ with rotational quantum number $J$ = 0 to 80.)  As shown by Neufeld \& Yuan (2008, Appendix), and contrary to earlier results obtained by Flower \& Gusdorf (2009), the assumption of steady-state is generally a good one in astrophysical environments, because the timescale to reach equilibrium is short compared to that on which the physical parameters vary.  Optical depth effects were treated using an escape probability method.  The only external source of radiation was assumed to be the cosmic microwave background (CMB) at $z=0$ (Fixsen et al.\ 2009), and molecular hydrogen was assumed to be the dominant collision partner, with the H$_2$ ortho-to-para ratio assumed to have its LTE value.  With these assumptions, the level populations depend on three physical parameters: the gas temperature, $T$, the H$_2$ density, $n({\rm H}_2)$, and a column density parameter, ${\tilde N} ({\rm CO})$, defined by Neufeld \& Melnick (1993).  For a medium with a large velocity gradient, $dv_z/dz$, in a single direction, ${\hat z}$,  ${\tilde N} ({\rm CO})$ is given by  $N ({\rm CO})\vert dv_z/dz \vert^{-1} $.  It is for this geometry that the calculation is actually performed, but Neufeld \& Melnick (1993) have discussed how the expression for ${\tilde N} ({\rm CO})$ may be scaled to obtain results for different cases (including approximate results for media in which the large velocity gradient limit does not apply.)

\subsection{Molecular data}

I obtained the CO rotational energies from the spectroscopic constants given by Nolt et al.\ (1987), and adopted the spontaneous radiative rates obtained by Goorvitch (1994).  Except at high densities sufficient to drive the the level populations to LTE, rate coefficients for the excitation of CO in collisions with H$_2$ are of critical importance in the interpretation of the CO line intensities.  Very recently, new estimates of these rate coefficients have been presented by Yang et al.\ (2010; hereafter Y10), who used the quantum-mechanical coupled-channels and coupled-states methods to compute state-to-state cross-sections, given a recent and highly-accurate potential energy surface obtained by Jankowski \& Szalewicz (2005).  The Y10 rate coefficients covered all transitions with upper states of rotational quantum number $J \le 40$, for temperatures spanning the range 2 to 3000 K, and with both ortho- and para-H$_2$ as the assumed collision partner.  Even though this set of collisional rate coefficients is the most extensive of any yet calculated, it is not sufficient to cover the large range of CO rotational transitions available to {\it Herschel}/PACS, and thus must be extrapolated to include states with $J > 40$. The need for extrapolation is obvious for PACS transitions higher than $J = 40 - 39$, but is also important for lower transitions, since even states with $J \le 40$ may be significantly populated via a radiative cascade from states of higher $J$. 

I have performed an extrapolation using the general methodology adopted by Schoier et al.\ 2005; hereafter S05) and other previous authors, in which (1) the subset of rate coefficients for deexcitation to the ground rotational state, $\gamma(J_U,0)$, is extrapolated to states of higher $J_U$; and then (2) this subset is ``expanded'' using some expression based upon the infinite order sudden (IOS) approximation to obtain a complete set of rate coefficients, $\gamma(J_U,J_L).$  However, some differences in the details of my extrapolation method are as follows:

First, in extrapolating the set of $\gamma(J_U,0)$, I have adopted an expression of the form 

$$\gamma(J_U,0)= A(T) \exp[-C(T)(E_U/kT)^{1/2}], \eqno(2)$$
where $E_U$ is the energy of the upper state, $J_U$, and $A(T)$ and $C(T)$ are computed at each temperature from a simultaneous fit to the Y10 values for $J_U$ = 35 and 40.  This expression differs from that used by S05, who assumed a quadratic (rather than linear) dependence of $\ln\gamma(J_U,0)$ upon $J_U$.  A detailed examination of the results of the S05 extrapolation method indicate that it leads to unphysical behavior at large $J_U$, and thus I favored the more robust extrapolation expression given in equation (2).  The Y10 rate coefficient are shown in Figure 1 (squares) for two example temperatures, with the extrapolation obtained from equation (2) plotted as dashed curves.  Red and blue symbols and curves apply respectively to deexcitation in collisions with ortho and para-H$_2$.  The red and blue curves in Figure 2 shows the values obtained for $C$, as a function of temperature, both for ortho- (red) and para-H$_2$ (blue).
  
Second, I have used a different expression to obtain the complete set of rate coefficients, $\gamma(J_U,J_L)$ from those obtained for $J_L=0$.  In the limit of large temperature, $kT \gg E_U - E_L$,
Goldflam, Green \& Kouri (1977, hereafter GGK) and Varshalovich \& Khersonskii (1977) independently used the infinite order sudden (IOS) approximation to obtain the following relation
$$\gamma(J_U,J_L) = 2 (J_L+1) \sum_{k=J_U-J_L}^{J_U+J_L} (2k+1) \left( \begin{array}{ccc} J_U & J_L & k \\ 0 & 0 & 0 \end{array} \right)^2  \gamma(J_U,0),\eqno(3)$$ where
$$\left( \begin{array}{ccc} J_ U & J_U & k \\ 0 & 0 & 0 \end{array} \right)$$ 
is a Wigner 3-j symbol.  To correct deficiencies in the IOS approximation that become increasingly important as $kT / (E_U - E_L) $ decreases, McKee et al.\ (1982) and S05 included a correction factor inside the summation on the right-hand-side of equation (3), derived from the work of de Pristo et al.\ (1979).    In the present work, by contrast,  I have multiplied the entire right-hand side by a phenomenological correction factor of the form $$G(J_U,J_L)= 1 + c_1\,\biggl[{J_L \over (T/{\rm K})^{1/2}}\biggr]^{c_2}\,\biggl[{J_U - J_L \over (T/{\rm K})^{1/2}}\biggr]^{c_3}. \eqno(4)$$
By optimizing the fit to rate coefficients that {\it are} available from Y10, I determined that the optimum values for the constants, $c_1$, $c_2$ and $c_3$ are 0.82, 2.20, and 1.22 respectively.
Figure 3 shows the relative performance of these various methods used to ``expand" the set of rate coefficients for cases ($J_L < J_U \le 40$) in which the Y10 results are available as a test.  In each panel, the vertical axis represents the initial rotational quantum number, $J_U$, and the horizontal axis shows the final quantum number, $J_L$, with the color of each tile representing the ratio of the rate coefficient computed using one of the three methods described above to the actual value obtained by Y10: these are referred to as ``GGK" (the use of equation 3 without correction for deficiencies in the IOS approximation), ``S05" (the correction introduced by de Pristo 1979 and used by McKee et al. 1982 and S05), and ``present work" (the correction obtained by multiplying the right-hand-side of equation 3 by the factor  
$G(J_U,J_L)$ given in equation 4).  Clearly, at high temperature ($T  = 1000\,\rm K$), equation (3) performs well for all transitions, but at a lower temperature ($T=100$~K), the IOS approximation systematically underestimates the rate coefficient except when $J_L$ is close to zero or $J_U$.  The correction used by S05 improves the fit, but  the correction factor given in equation (4) clearly provides the best performance (even though it lacks a clear physical basis.)

Figure 4 shows example rate coefficients adopted in the present work, with the vertical axis in each panel representing the initial rotational quantum number, $J_U$, the horizontal axis the final quantum number, $J_L$, and the color of each tile representing the rate coefficient.  Tiles below the horizontal white line show the values given by Y10, while those above the horizontal white line show results of the extrapolation method adopted here\footnote{The complete set of extrapolated rate coefficients is available from the author upon request, together with solutions to the equations of statistical equilibrium for the level populations; the latter were obtained on a fine grid of temperatures and densities, and are available as a IDL save set or a FITS file}.

\section{The CO rotational diagram for an optically-thin and isothermal medium}

The optically-thin and isothermal medium represents the simplest case for which a rotational diagram can be determined: the fractional level populations, $f_J = N_J/N({\rm CO})$, are simply computed from the equations for statistical equilibrium and then presented on a log-linear plot.
Figure 5 shows the rotational diagrams predicted for such a medium, assumed in this case to be at a temperature of 1000~K. The thick curves apply to H$_2$ densities of 10$^4$ (red), 10$^5$ (green), 10$^6$ (blue), 10$^7$ (magenta), and 10$^8$~cm$^{-3}$ (black), with the thinner black curves applying to intermediate densities spaced by 0.2~dex.  The same results are presented in the upper panel of Figure 6, but now with the values of $(f_J/g_J)$ normalized relative to that obtained for the lowest rotational state accessible with {\it Herschel}/PACS ({\it viz}.\ $J=13$).  Horizontal bars indicate the range of transitions accessible in four separate bands of the PACS spectrometer -- denoted Long R1, Short R1, B2B and B2A -- for the case in which the observations are acquired in the commonly-used ``SED scan" mode (Poglitsch et al.\ 2010).  The lower panel in Figure 6 shows the rotational temperature as defined in equation (1): $T_{\rm rot} \equiv -(k\,{\rm dln}\,[N_J/g_J]/dE_J)^{-1}$. 

In the high density limit, the level populations approach LTE, the rotational diagram is well-approximated by a straight line, and $T_{\rm rot}$ approaches the kinetic temperature, $T$.  As the density decreases, the fractional level populations for transitions in the PACS wavelength range drop, with the largest reductions occurring for the highest-lying transitions (which have the largest spontaneous radiative decay rates.)  Figure 6 shows that for densities in the range 10$^{6.8}$ to 10$^8\rm \, cm^{-3}$, subthermal excitation leads to a {\it negatively}-curved rotational diagram in which  $T_{\rm rot}$ is a monotonically {\it decreasing} function of $E_J$.  However, at yet lower densities, departures from LTE can introduce a {\it positive} curvature in the rotational diagram (i.e.\ the behavior that is typically observed in nature).  For densities $\le 10^{4.8}\,\rm cm^{-3}$,  $T_{\rm rot}$ is a {\it monotonically increasing} function of $E_J$ for the set of transitions accessible to PACS.

\subsection{Results in the low density limit}

Rotational diagrams for the low density limit are shown in Figure 7, with the density fixed at 10$^3\,\rm cm^{-3}$ and for temperatures of $10^2$ (solid red curve), 10$^{2.5}$ (green), $10^3$ (blue), and 10$^{3.5}$ K (magenta).  The behavior can be understood analytically by means of the following analysis.  At low densities, most CO molecules are in the lowest two rotational states (the $J=1$ state being populated by the absorption of CMB radiation).  Any state with $J \ge 13 $ (i.e.\ in the range observable by PACS) is {\it depopulated} primarily as a result of spontaneous radiative decay, at a rate that is roughly proportional to $J^3$, and is {\it populated} primarily by collisional excitation to states with $J^\prime \ge J$ (followed by a radiative cascade except in the case $J^\prime = J$.)  The excitation rate from $J=0$ or 1 to $J^\prime$ is roughly proportional to $g_{J^\prime} \exp(-x_{J^\prime}^2-Cx_{J^\prime})$, where $x_{J^\prime} \equiv (E_{J^\prime}/kT)^{1/2}$.  This last proportionality follows from the roughly linear behavior of the curves plotted in Figure 1.  In the limit $J^\prime \gg 1$, we also have $E_{J^\prime} \propto J^{\prime 2}$ and $g_{J^\prime} \propto J^\prime$.  With these approximations, the equilbrium level population therefore varies with $J$ in accord with

$${f_J \over g_J}  \propto
{1 \over x_J^4}\int^\infty_{x_J} x e^{-x^2-Cx} dx \\
= {1\over 4x_J^4} \biggl[e^{-x_J^2-Cx_J} (2 -  \pi^{1/2}C e^{(x_J+C/2)^2} {\rm erfc}\, (x_J+\onehalf C)) \biggr], \eqno(5)
$$
where ${\rm erfc}(z)$ is the complementary error function and a sum over $J^\prime$ has been approximated by an integral.  The corresponding rotational temperature is given by 
$$kT_{\rm rot} \equiv -(\,{\rm dln}\,[f_J/g_J]/dE_J)^{-1} = 2x_JE_J \biggl[{4 \over x_J} + {4x_J \over 2 -  \pi^{1/2}C e^{(x_J+C/2)^2} {\rm erfc}\, (x_J+\onehalf C)} \biggr]^{-1}.\eqno(6)$$

I have treated the quantity $C$ appearing in equations (5) and (6) as an adjustable parameter, which I have optimized for each temperature to achieve the best fit to the numerical results.  The best-fit values for $C$ are plotted in Figure 2 (black curve), as a function of temperature, and roughly follow\footnote{The black curves are systematically lower, because the curves plotted in Figure 1 show a slightly negative curvature: i.e.\ the best fit value of $C$ for $J \le 20$ is somewhat smaller than that ($J= 35  - 40$) adopted for the purpose of extrapolation.} the analogous values used to extrapolate the rate coefficients to $J_U \ge 40$.   The dashed curves in Figure 7 present the analytic results given in equations (5) and (6), obtained using these best-fit values of $C$; they evidently yield excellent approximations to the numerical results.

The quantity $C$ shows only weak dependence on temperature (Figure 2), and adopting a constant value of 5 yields a reasonably good approximation to all results obtained for $T \ge 100$~K.  Because the argument of the complementary error function in equation (6) is always $\ge 2.5$, a further simplifying approximation is possible: for large $z$, erfc($z$) may be approximated by $\exp(-z^2)/(\pi^{1/2}z)$, the error in this approximation being $\le 7\%$ for all $z \ge 2.5$.  This additional approximation permits equation (6) to be rewritten in the form
$$kT_{\rm rot} \simeq {x_J^2 kT \over  x_J^2  +\onehalf Cx_J + 2 } \simeq {E_J \over E_J/kT + 2.5\,(E_J/kT)^{1/2} + 2}, \eqno(7)$$
which is plotted as the dotted lines in Figure 7 (lower panel).

\subsection{Comparison with PACS data}

To facilitate comparison with PACS observations, I have computed the average slopes in the rotational diagrams expected for each of the 4 PACS spectrometer bands: Long R1 (covering 140 -- 220~$\mu$m), Short R1 (102 -- 146~$\mu$m), B2B (70-105~$\mu$m), and B2A (51-73~$\mu$m).  These average slopes lead to average rotational temperatures $T_{\rm LR1}$, $T_{\rm LR2}$, $T_{\rm B2B}$ and $T_{\rm B2A}$ that may be conveniently compared with fits to the observational data.  Clearly, if measurable, the combination of the longest and shortest wavelength spectral bands yields the maximum ``leverage" on the physical parameters.  Figure 8 shows contour plots of $T_{\rm LR1}$ (red), $T_{\rm B2B}$ (green) and $T_{\rm B2A}$ (blue) in the ${\rm log}\,T - {\rm log}\,n({\rm H}_2)$ plane.  While the intent of this paper is not to provide detailed modeling of individual sources, I note that the rotational diagrams observed with PACS toward embedded protostars have typically been fit as a sum of two components in LTE (e.g.\ van Kempen et al.\ 2010a,b; Fich et al.\ 2010; Herczeg et al.\ 2011): a ``warm" component at $T_{\rm warm} \sim 250 - 400$~K, and a ``hot" component at $T_{\rm hot} \sim 800 - 1400$~K.  In sources -- such as NGC1333 IRAS4B (Herzceg et al.\ 2011) -- where the observed CO emission extends to band B2A, and where $T_{\rm hot}$ lies close to the lower end of the latter range (i.e.\ 800~K), the results in Figure 8 suggest that a single temperature component can explain the rotational diagram; a gas density $\sim 10^4 - 10^5$~cm$^{-3}$ and a temperature $\sim 3000 - 5000$~K (i.e\ close to the maximum value at which CO can survive) are required to account for the observed rotational diagram.  On the other hand, sources with larger fitted values of $T_{\rm hot}$ (e.g. DK Cha, observed by van Kempen et al.\ 2010a) 
have rotational diagrams with  larger (more positive) curvature than can be accounted for by a single temperature, non-LTE component.

\subsection{Optical depth effects}

All the calculations reported in \S3.1 and 3.2 above apply to optically-thin CO emission from an isothermal medium.  CO has a dipole moment (0.11~Debye) that is much smaller than that of most other heteronuclear molecules, with the result that its high-$J$ rotational transitions are usually optically-thin despite the large abundance of CO.  Optical depth effects are examined\footnote{One subtlety of the optically-thick case is that CO rotational diagrams are typically computed {\it under the assumption that every emission line is optically-thin.}  In order to make a comparison with results presented in the literature, it is therefore necessary to obtain predictions for the level populations that {\it would} have been inferred {\it under that assumption}; in effect, the quantity plotted on the vertical axis is not exactly $N_J/g_J$ but, in reality, $4 \pi I_{UL} / [A_{UL} g_U (E_U - E_L]) $, where $I_{UL}$ is the integrated line flux.  The rotational temperatures plotted in Figure 9 were computed accordingly.} in Figure 9, where contour plots for $T_{\rm LR1}$ (red) and $T_{\rm B2A}$ (blue) are shown for four values of the optical depth parameter, $\tilde N$(CO) (see \S2.1 for the definition): 10$^{15}$, 10$^{16}$, 10$^{17}$ and 10$^{18}$ cm$^{-2}$ per km/s.  For a CO/H$_2$ ratio of 10$^{-4}$, non-dissociative molecular shocks are expected to yield $\tilde N$ $\sim 10^{16}$ cm$^{-2}$ per km/s (e.g. Neufeld \& Kaufman 1993); thus the results shown in Figure 9 suggest that optical depth effects are likely to be negligible for the parameters typically encountered in warm shock-heated gas.  

\section{The CO rotational diagram for a continuous distribution of gas temperatures}

As shown above, even an isothermal medium can give rise to a CO rotational diagram with positive curvature -- provided the gas density is sufficiently small -- and may be successful in accounting for the CO rotational diagrams observed in some sources.  Nevertheless, it is of interest to consider also the case where a distribution of gas temperatures is present within the telescope beam or along the line-of-sight.  Such an admixture of temperatures might well occur if multiple shocks of varying velocities are present within the beam, or if shocks and photodissociation regions (PDRs) are both present (e.g. Visser et al.\ 2011).

Evidence for multiple warm gas components of different temperatures has been inferred from H$_2$ rotational diagrams observed with {\it Spitzer}, which show a positive curvature that cannot be accounted for by an isothermal medium (e.g.\ Neufeld \& Yuan 2008).  In addition, the observed H$_2$ ortho-to-para (OPR) ratio is typically considerably smaller than the equilibrium value of 3 expected in warm molecular gas, with a systematic tendency for the OPR inferred from observations of the lowest pure rotational transitions to be furthest from equilibrium; this behavior finds a natural explanation (e.g. Neufeld et al.\ 2006) if an admixture of gas temperatures is present, because the rate at which the OPR is expected to reach its equilibrium value is a strongly increasing function of temperature.  Previous studies (e.g.\ Neufeld \& Yuan 2008) have been successful in matching the observed H$_2$ rotational diagrams by positing a power-law temperature distribution, $dN/dT \propto T^{-b} db$, over some range of temperatures, $[T_{\rm min},T_{\rm max}]$.  The power-law index, $b$, typically has a best-fit value varying from one source to another within the range $\sim 2$ to 5 (Nisini et al.\ 2010, Yuan \& Neufeld 2011, Giannini et al.\ 2011).

In Figure 10, normalized rotational diagrams (analogous to those in Figure 6) are presented for a power-law distribution of gas temperatures, along with the corresponding rotational temperature.  These results were obtained for $T_{\rm min}=10$~K (although the exact value of $T_{\rm min}$ is immaterial provided it is $\ll E_J/k$), and for $T_{\rm max}=5000$~K; my choice of $T_{\rm max}$ is roughly the maximum temperature at which CO molecules can survive. The example results shown in Figure 10 are for a power-law index $b=3$ and the same range of densities plotted in Figure 6 (and with the results for H$_2$ densities of 10$^4$, 10$^5$, 10$^6$, 10$^7$, and 10$^8$~cm$^{-3}$ respectively appearing as thick curves in red, green, blue, magenta and black, as before).  
Positive curvature is predicted, both at low {\it and} high densities, with the rotational temperatures increasing monotonically with $E_J$ in every case.

\subsection{Results in the high density limit}

In Figure 11, results are presented for the high-density limit.   Here, the populations are assumed to be in LTE, and results are shown for power-law indices, $b$, of 1 (black), 2 (magenta), 3 (blue), 4 (green) and 5 (red).   An analytic treatment, related to that
introduced by Mauersberger et al.\ (1988; their Appendix A) to treat the excitation of interstellar ammonia, 
 can account exactly for the plotted results.  For a power-law distribution of gas temperatures, and where the level populations are in LTE, the total population in state $J$ obeys the proportionality
$${f_J \over g_J} \propto \int_{T_{\rm min}}^{T_{\rm max}} {e^{-E_J/kT} \over Z_{\rm rot}(T)} T^{-b} dT \propto {\Gamma(b,z_1) - \Gamma(b,z_2) \over E_J^b}  \eqno(8)  $$
where $Z_{\rm rot}(T) \propto T$ is the rotational partition function, $z_1 \equiv E_J/kT_{\rm min}$, $z_2 \equiv E_J/kT_{\rm max}$ and $\Gamma(b,z) \equiv \int_z^\infty t^{b-1}e^{-t}dt$ is the upper incomplete gamma function.   Some further calculus and algebra yields the rotational temperature in the form
$$kT_{\rm rot} \equiv -(\,{\rm dln}\,[f_J/g_J]/dE_J)^{-1} = {\Gamma(b,z_1) - \Gamma(b,z_2) \over \Gamma(b+1,z_1) - \Gamma(b+1,z_2)}\,E_J \eqno(9)$$ 
Dashed black curves show the results obtained from the analytic treatment, which are indistiguishable from the numerical results.  For all the calculations performed here, $E_J \gg T_{\rm min} = 10$~K, and equation (9) becomes 
$kT_{\rm rot} = E_J \Gamma(z_1,b) / \Gamma(z_1,b+1)$.
If, in addition, $E_J \ll T_{\rm max} = 5000$~K, this expression reduces further to $kT_{\rm rot} = E_J/b$.  The behavior in this limit explains the linear dependence of $T_{\rm rot}$ on $E_J$ in the left part of the lower panel.

\subsection{Comparison with PACS data}

Figure 12 is analogous to Figure 8, now showing contours of $T_{\rm LR1}$ (red), $T_{\rm B2B}$ (green) and $T_{\rm B2A}$ (blue) as a function of the power-law index, $b$, and the density.  By comparison with the isothermal case, a medium with a distribution of gas temperatures can yield rotational diagrams with larger positive curvature.  This is manifested by larger ratios of $T_{\rm B2A}$ to $T_{\rm LR1}$, which provide a better fit to observational results obtained in some sources (e.g.\ DK Cha, observed by van Kempen et al.\ 2010a).  In Figure 13, values of $T_{\rm B2A}$/$T_{\rm LR1}$ are shown for the two cases; the upper panel applies to a medium at a single temperature, while the lower panel applies to a medium with a power-law distribution of temperatures.   At low density, an isothermal medium can result in $T_{\rm B2A}$/$T_{\rm LR1}$ values as large as $\sim 4$, but values larger than 4 require an admixture of gas temperatures.  

\section{Summary}

1.  With the use of recent collisional rate coefficients computed by Y10, and an extrapolation method discussed in \S2.1, I have computed the far-infrared carbon monoxide emissions expected from warm astrophysical media.  

2.  For the case of an isothermal medium, results have been obtained for a wide range of gas temperatures (100 - 5000~K) and H$_2$ densities ($10^3 - 10^9 \, \rm cm^{-3}$), and presented in the form of rotational diagrams.  For rotational transitions in the wavelength range accessible to {\it Herschel}/PACS, such diagrams are nearly linear when $n({\rm H}_2)\ge 10^8$~$\rm cm^{-3}$.  When $n({\rm H}_2) \sim 10^{6.8} - 10^8$~$\rm cm^{-3}$, they exhibit significant negative curvature, whereas when $n({\rm H}_2) \le 10^{4.8}$~$\rm cm^{-3}$, the curvature is uniformly positive throughout the PACS-accessible range.   (In the intermediate density range, $n({\rm H}_2) \sim 10^{4.8} - 10^{6.8}$~$\rm cm^{-3}$, there is an inflection point where the curvature changes sign.)  Thus, the observation of a positively-curved CO rotational diagram does not necessarily require the presence of multiple temperature components.  

3.  For the isothermal case, an analytic treatment accounts well for the behavior in the low-density limit, providing simple expressions for the CO level populations and rotational temperatures.

4.  The average rotational temperatures expected for four PACS spectroscopic bands are presented as contours in the plane of density and temperature.  For some sources observed with {\it Herschel}/PACS, the CO rotational diagrams show a modest positive curvature that can be explained by a single isothermal component.  Typically, the required physical parameters are densities in the $10^4 - 10^5$~cm$^{-3}$ range and temperatures close to the maximum at which CO can survive.  Other sources exhibit rotational diagrams with more curvature than can be accounted for by a single temperature component.  
 
5.  Optical depth effects are found to be very modest, given the typical CO column densities in warm astrophysical media.
 
6.  For the case of a medium with a power-law distribution of gas temperatures, $dN/dT \propto T^{-b}$,  results have been obtained for H$_2$ densities $10^3 - 10^9 \, \rm cm^{-3}$ and power-law indices, $b$, in the range 1 to 5.  An analytic expression is given for the CO level populations and rotational temperatures in the limit of high density. The average rotational temperatures expected for four PACS spectroscopic bands are presented as contours in the plane of density and temperature.

7.  A medium with a distribution of gas temperatures can account for a CO rotational diagram that is more positively curved than any resulting from an isothermal medium.

\begin{acknowledgments}

It is a pleasure to acknowledge useful discussions with  M.~Puravankara, F.~van~der~Tak,  R.~Visser, and D.~Watson.  Support for this work was provided by NASA through an award issued by JPL/Caltech.

\end{acknowledgments}

\vfill\eject
\begin{figure}
\includegraphics[scale=0.75,angle=0]{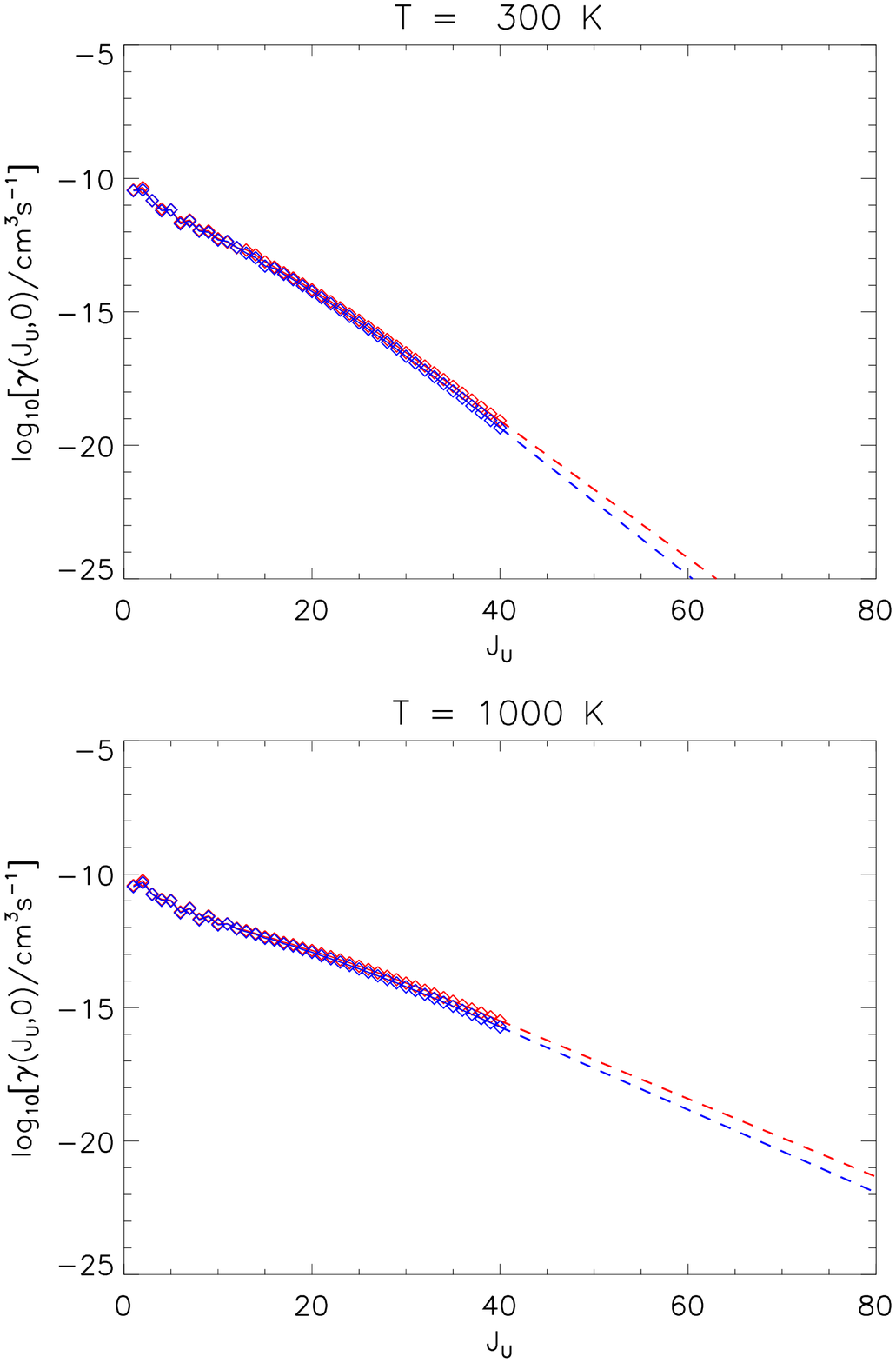}

\noindent{Fig.\ 1 -- Rate coefficients for collisional deexcitation from $J = J_U \rightarrow 0$ at temperatures of 300~K (top panel) and 1000~K (bottom panel). Squares: results from Y10 for collisions with ortho-H$_2$ (red) and para-H$_2$ (blue).  Dashed lines: extrapolation adopted in the present work (equation 2).}
\end{figure}

\vfill\eject
\begin{figure}
\includegraphics[scale=1.0,angle=0]{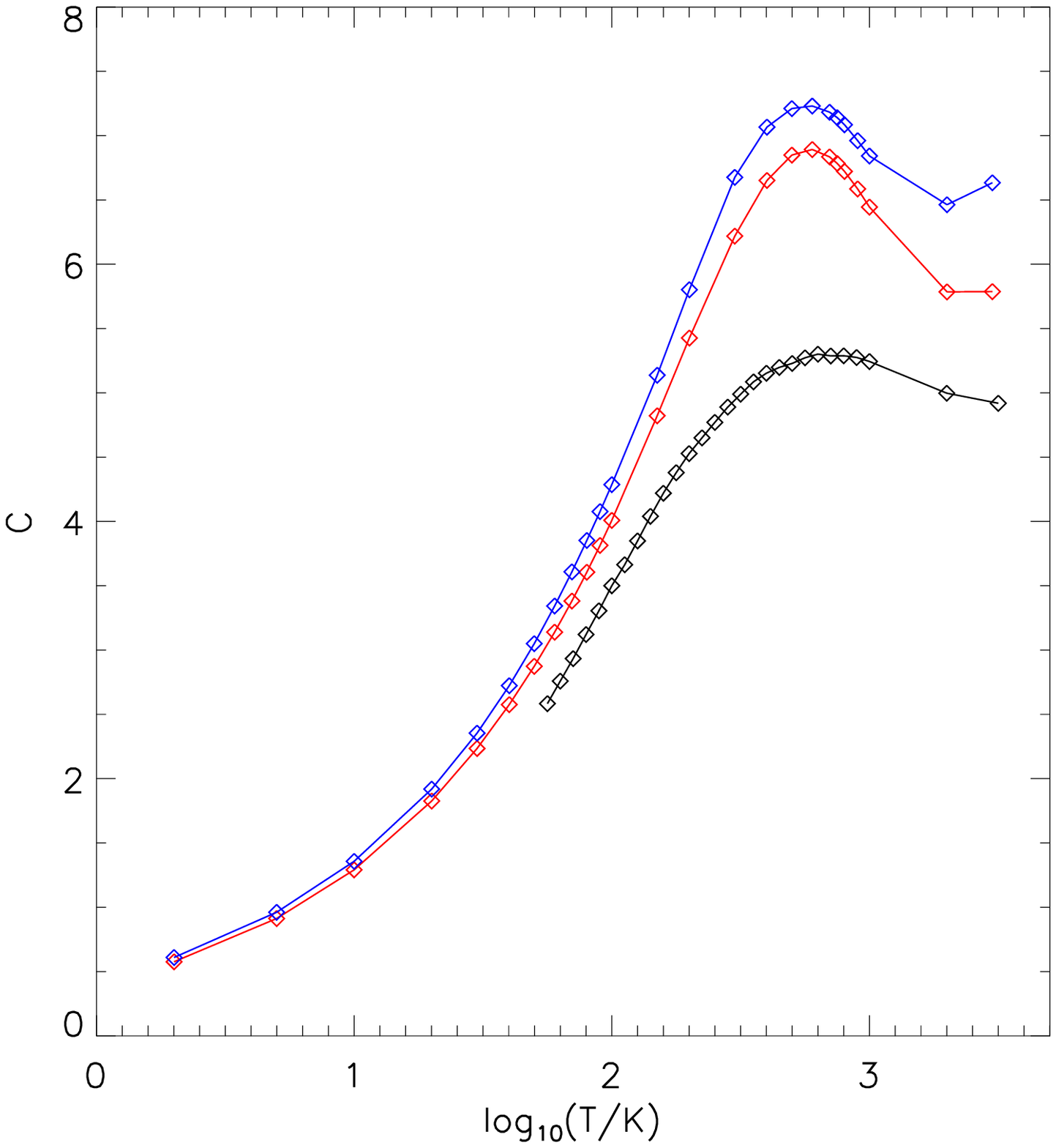}

\noindent{Fig.\ 2 -- Values adopted for $C(T)$, when fitting the $\gamma(J_U,0)$ by the expression $A(T) \exp[-C(T)(E_U/kT)^{1/2}]$}.  Red squares: values adopted in extrapolating the rate coefficients for deexcitation by ortho-H$_2$.  Blue squares: values adopted in extrapolating the rate coefficients for deexcitation by para-H$_2$.  Black squares: values that yield the best fit to the rotational diagram (as approximated by equation (5) in \S3.1).

\end{figure}

\vfill\eject
\begin{figure}
\includegraphics[scale=0.6,angle=0]{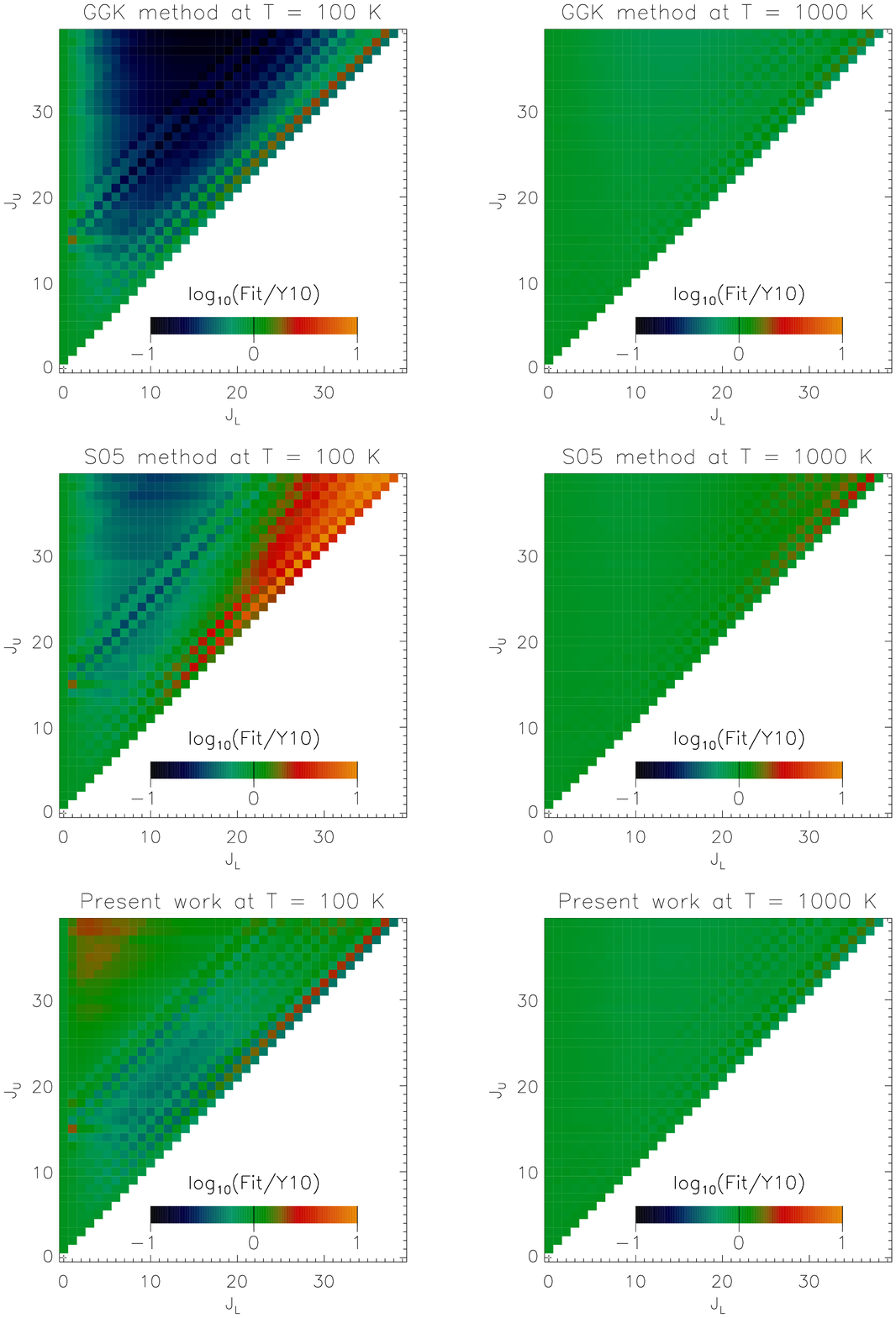}

\noindent{Fig.\ 3 -- Relative performance of various methods (see text) used to ``expand" the set of rate coefficients for cases ($J_L < J_U \le 40$) in which the Y10 results are available as a test.  In each panel, the vertical axis represents the initial rotational quantum number, $J_U$, and the horizontal axis shows the final quantum number, $J_L$.   The color of each tile indicates the ratio of the rate coefficient computed using each method to the actual value obtained by Y10. }
\end{figure}

\vfill\eject
\begin{figure}
\includegraphics[scale=0.6,angle=0]{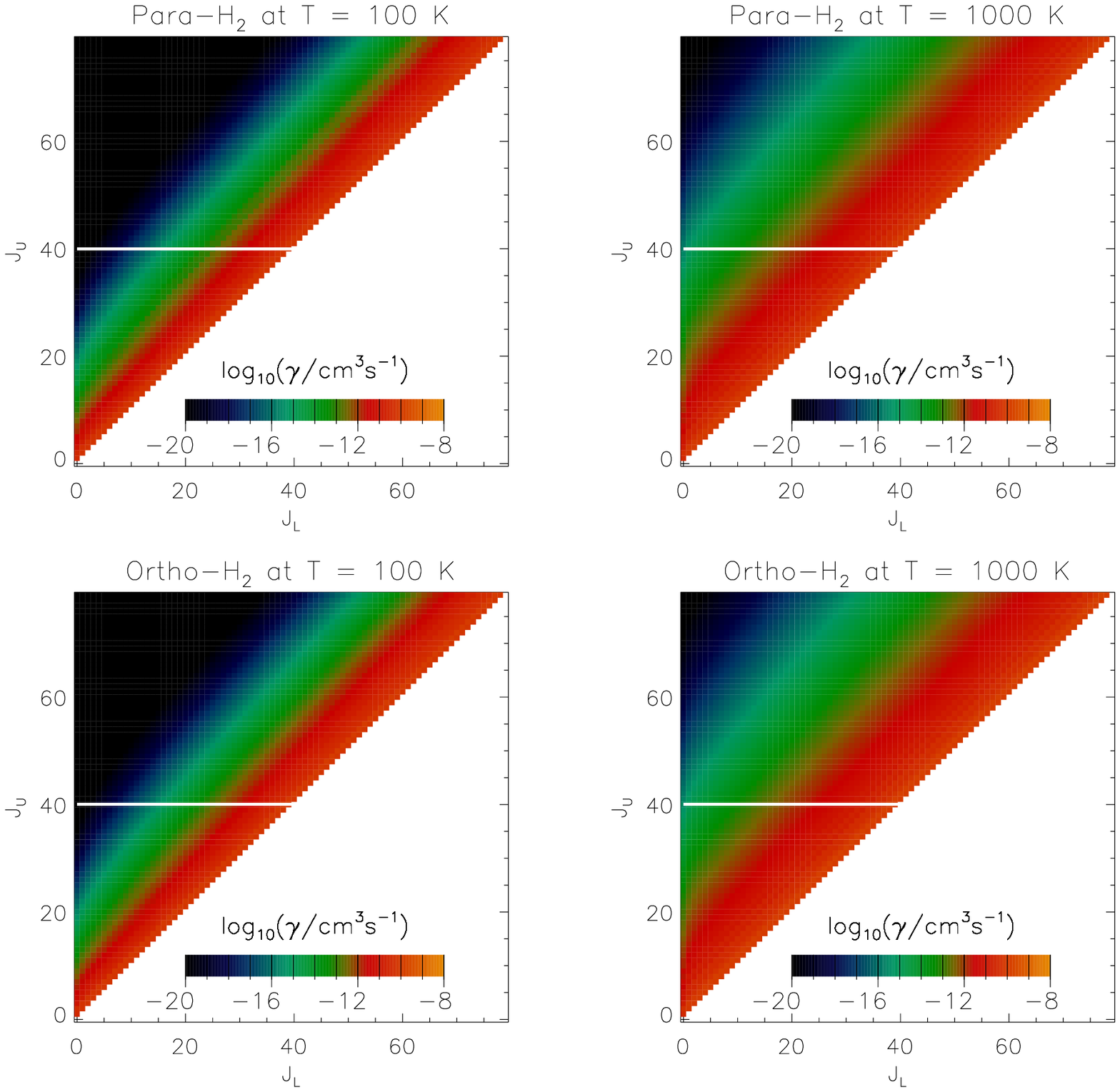}

\noindent{Fig.\ 4 -- Rate coefficients for collisional deexcitation of CO by H$_2$, with the vertical axis in each panel representing the initial rotational quantum number, $J_U$, the horizontal axis the final quantum number, $J_L$, and the color of each tile representing the rate coefficient.  Tiles below the horizontal white line: values given by Y10.  Tiles above the horizontal white line: extrapolation described in \S2.1}
\end{figure}

\vfill\eject
\begin{figure}
\includegraphics[scale=1.0,angle=0]{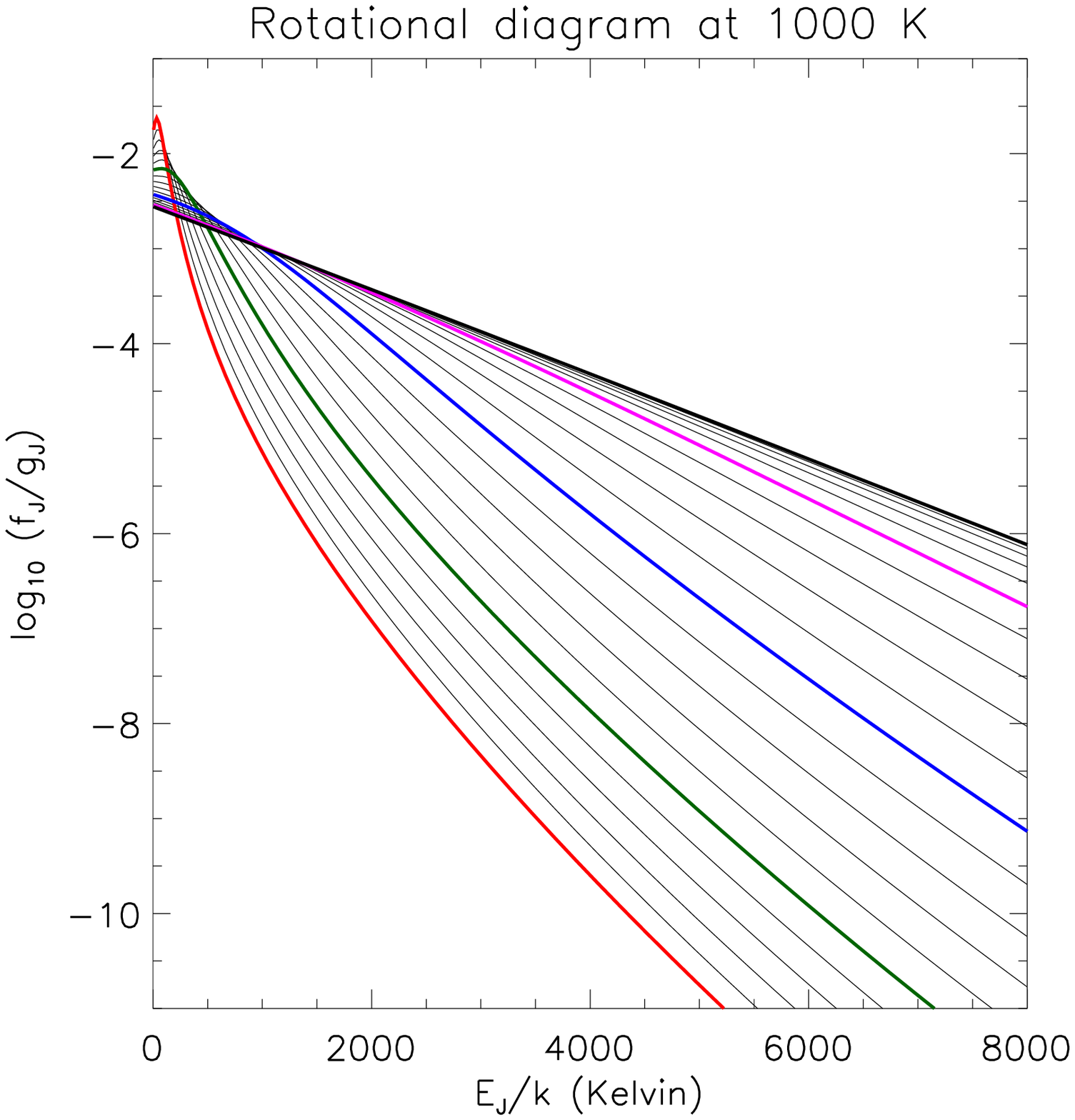}

\noindent{Fig.\ 5 -- Rotational diagrams predicted for an optically-thin, isothermal medium at a temperature of 1000~K. The thick curves apply to H$_2$ densities of 10$^4$ (red), 10$^5$ (green), 10$^6$ (blue), 10$^7$ (magenta), and 10$^8$~cm$^{-3}$ (black), with the thinner black curves applying to intermediate densities spaced by 0.2~dex.}
\end{figure}

\vfill\eject
\begin{figure}
\includegraphics[scale=0.7,angle=0]{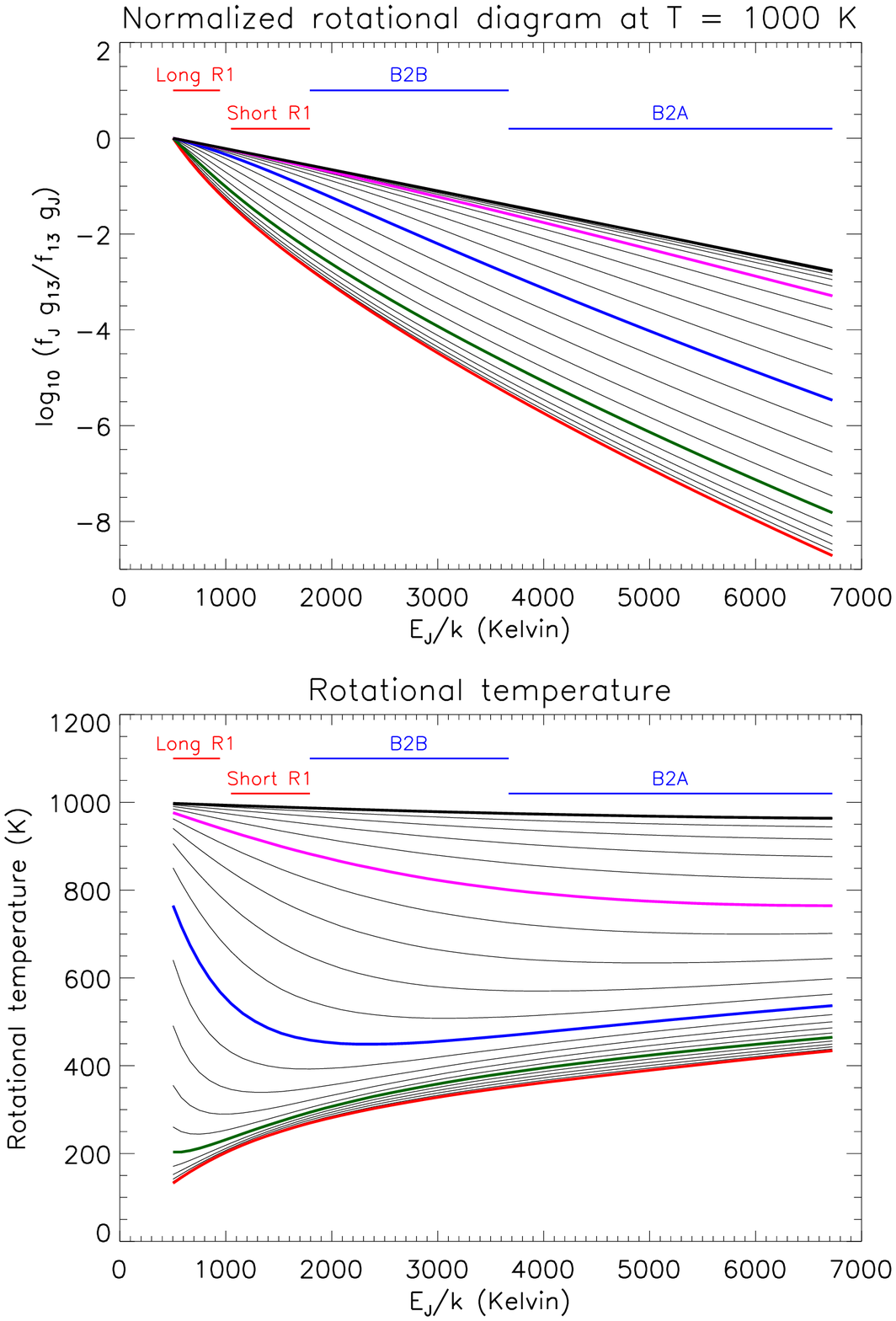}

\noindent{Fig.\ 6 -- Upper panel: same as Figure 5, but with the level populations normalized to those for $J=13$.  Bottom panel: corresponding rotational temperatures, $T_{\rm rot} \equiv -(k\,{\rm dln}\,[N_J/g_J]/dE_J)^{-1}$.  Horizontal red and blue bars indicate the range of rotational levels probed by observations in each of the four labeled PACS spectral bands.}
\end{figure}

\vfill\eject
\begin{figure}
\includegraphics[scale=0.7,angle=0]{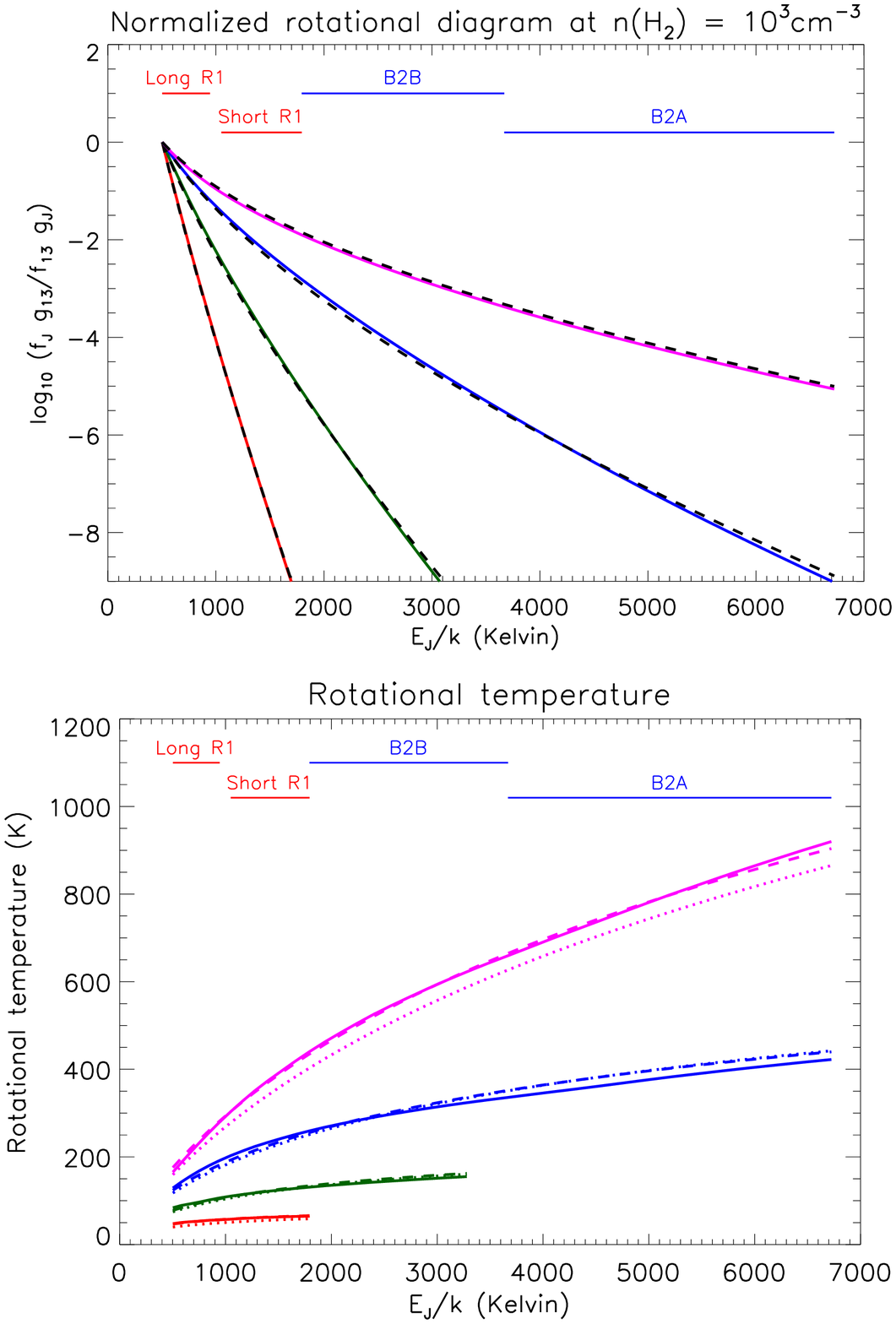}

\noindent{Fig.\ 7 -- Rotational diagrams (top panel) and rotational temperatures (bottom panel) predicted for an optically-thin, isothermal medium at an $\rm H_2$ density of $10^3\,{\rm cm}^{-3}$, with solid red, green, blue and magenta curves applying to temperatures of 10$^2$, 10$^{2.5}$, 10$^3$ and 10$^{3.5}$ K respectively, and the approximate analytic treatment (equation 5) shown as dashed curves.  The dotted curves were obtained with a further simplifying approximation (equation 7; see text).}

\end{figure}

\vfill\eject
\begin{figure}
\includegraphics[scale=0.7,angle=0]{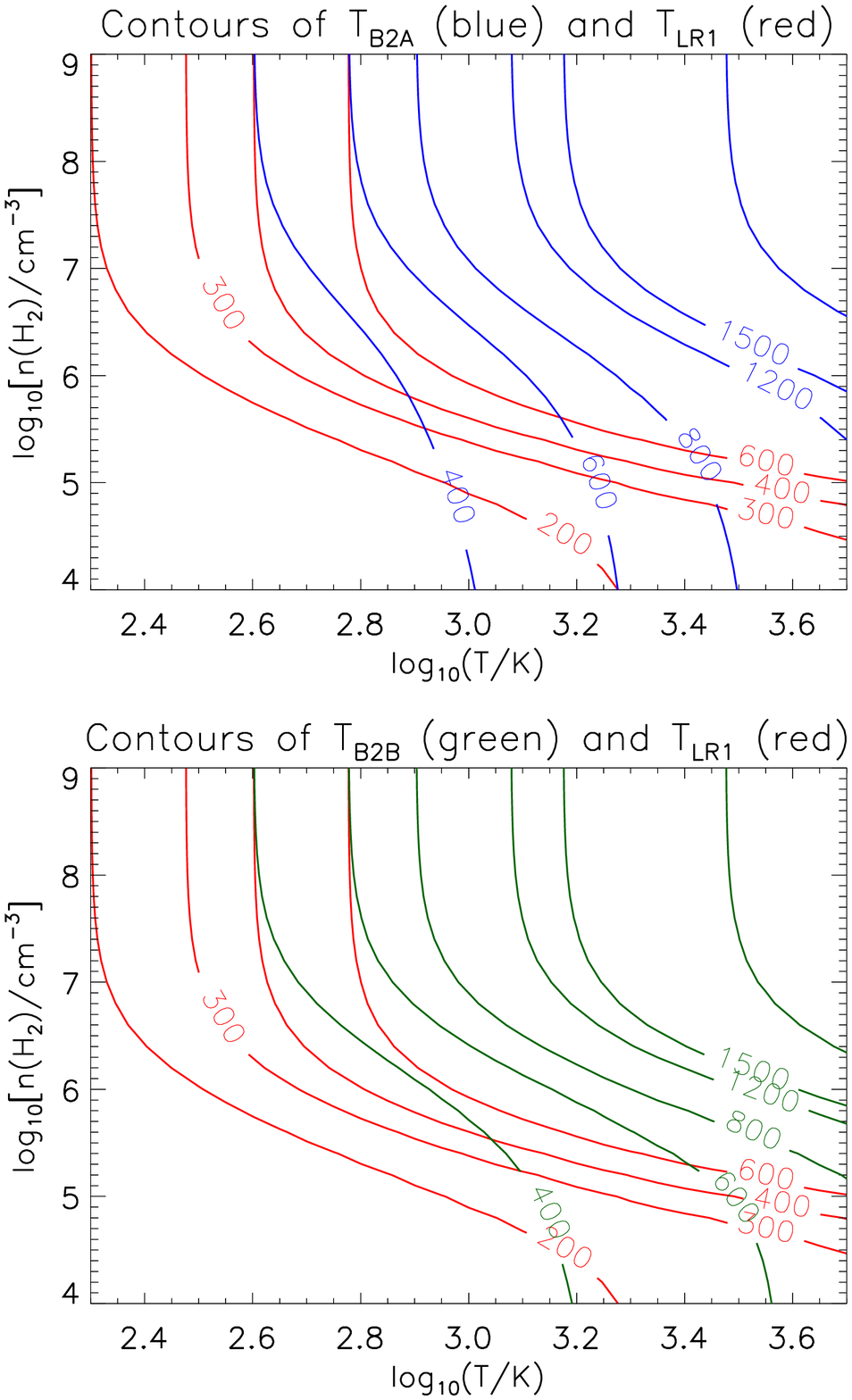}

\noindent{Fig.\ 8 -- Contour plots of the average rotational temperatures (in Kelvin) predicted for CO transitions in the Long R1 (140 -- 220~$\mu$m), B2B (70-105~$\mu$m), and B2A (51-73~$\mu$m) bands of {\it Herschel}/PACS, as a function of temperature and H$_2$ density. These are shown in red ($T_{\rm LR1}$), green ($T_{\rm B2B}$), and blue $T_{\rm B2A}$, respectively. Results apply in the optically-thin limit.}
\end{figure}

\vfill\eject
\begin{figure}
\includegraphics[scale=1.0,angle=0]{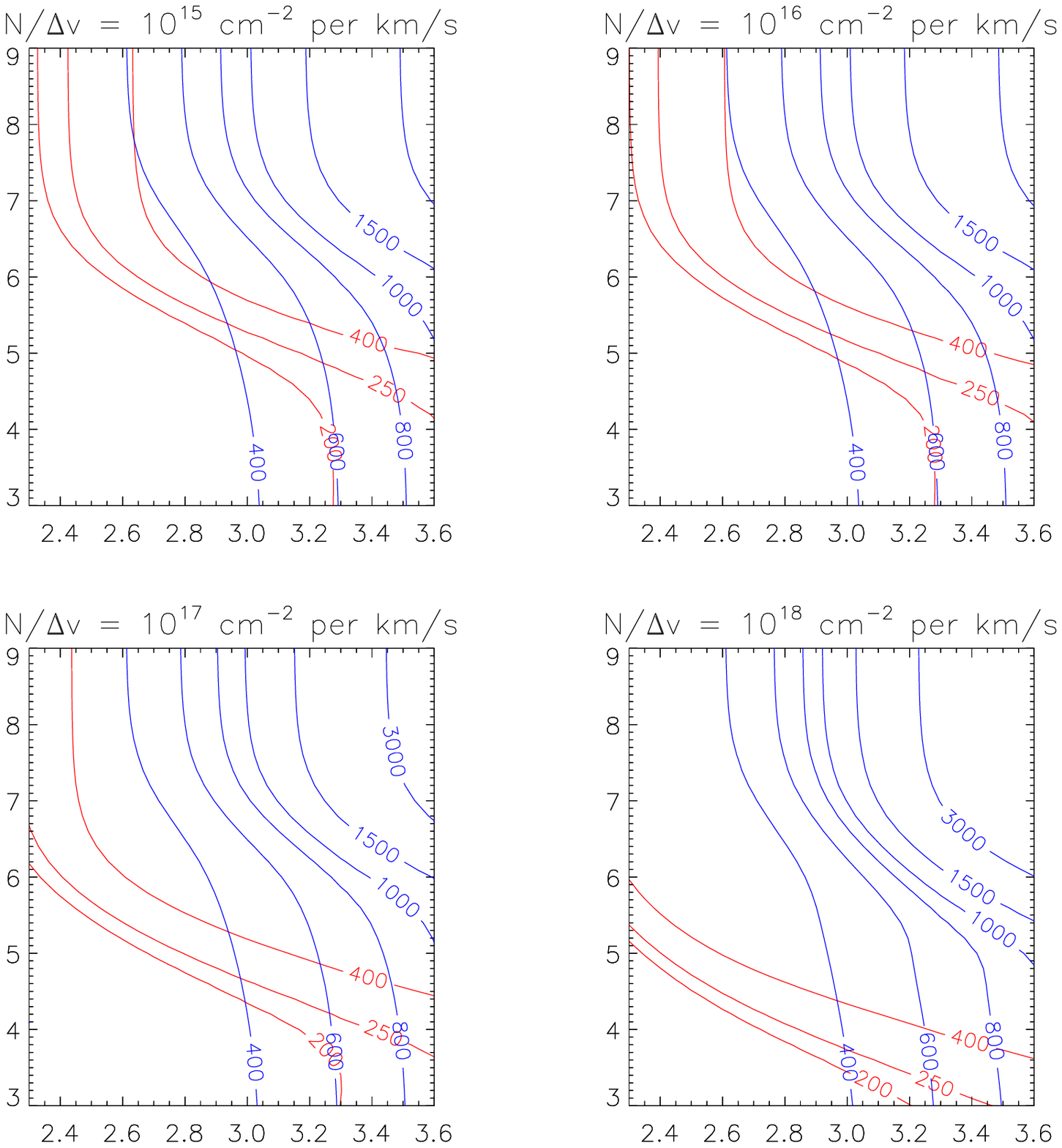}

\noindent{Fig.\ 9 -- Same as Figure 8, but for four different values of the optical depth parameter, $\tilde{N}$(CO) (see text).}
\end{figure}

\vfill\eject
\begin{figure}
\includegraphics[scale=0.65,angle=0]{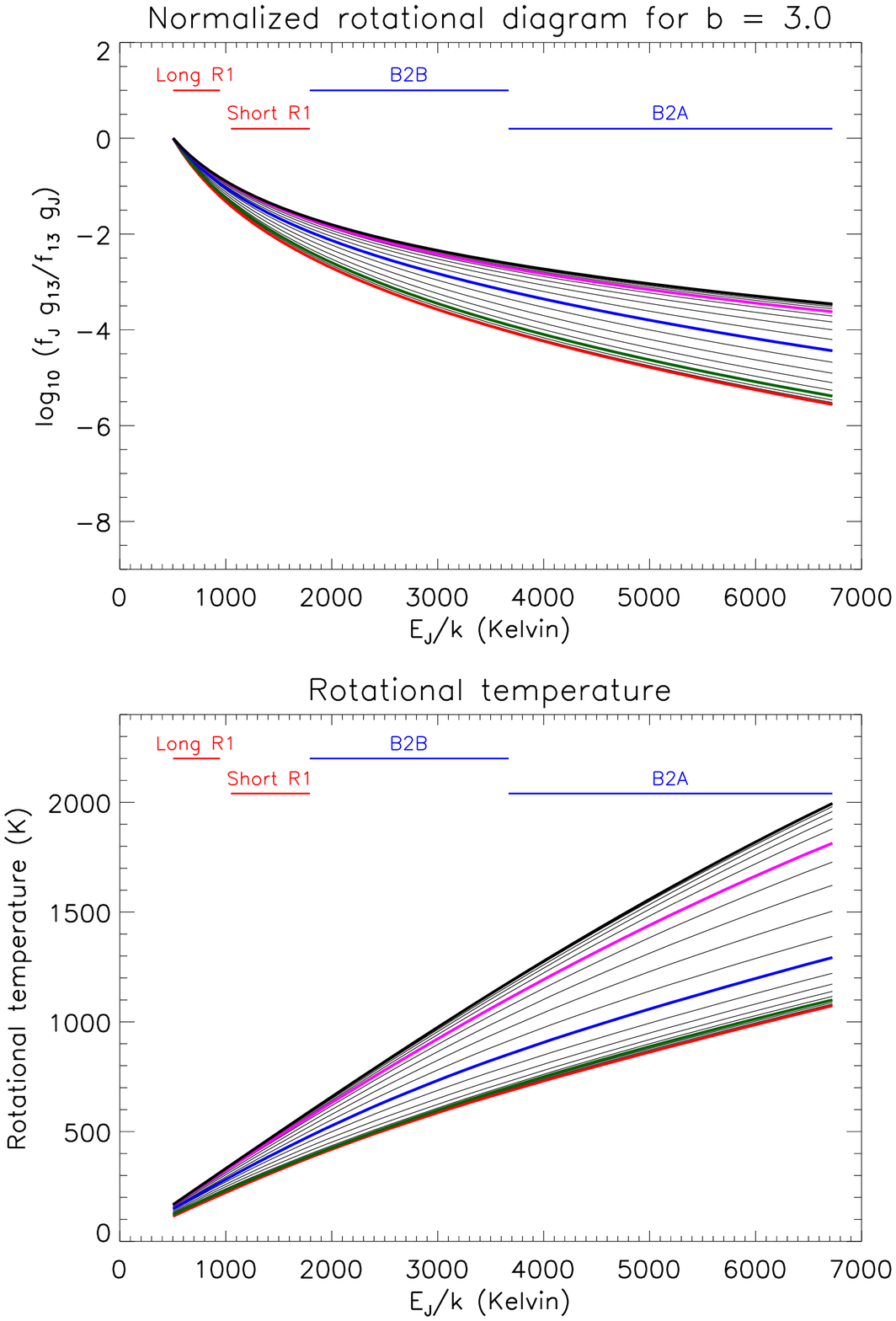}

\noindent{Fig.\ 10 -- Rotational diagrams, and rotational temperatures, $T_{\rm rot} \equiv -(k\,{\rm dln}\,[N_J/g_J]/dE_J)^{-1}$, predicted for an optically-thin medium with a power-law distribution of temperatures of the form $dN/dT = T^{-b} dT$ over the range 10 to 5000~K.
Thick curves apply to H$_2$ densities of 10$^4$ (red), 10$^5$ (green), 10$^6$ (blue), 10$^7$ (magenta), and 10$^8$~cm$^{-3}$ (black), with the thinner black curves applying to intermediate densities spaced by 0.2~dex.  The power-law index assumed here is $b=3.0$, and the level populations in the top panel are normalized with respect to those for $J=13$}

\end{figure}

\vfill\eject
\begin{figure}
\includegraphics[scale=0.7,angle=0]{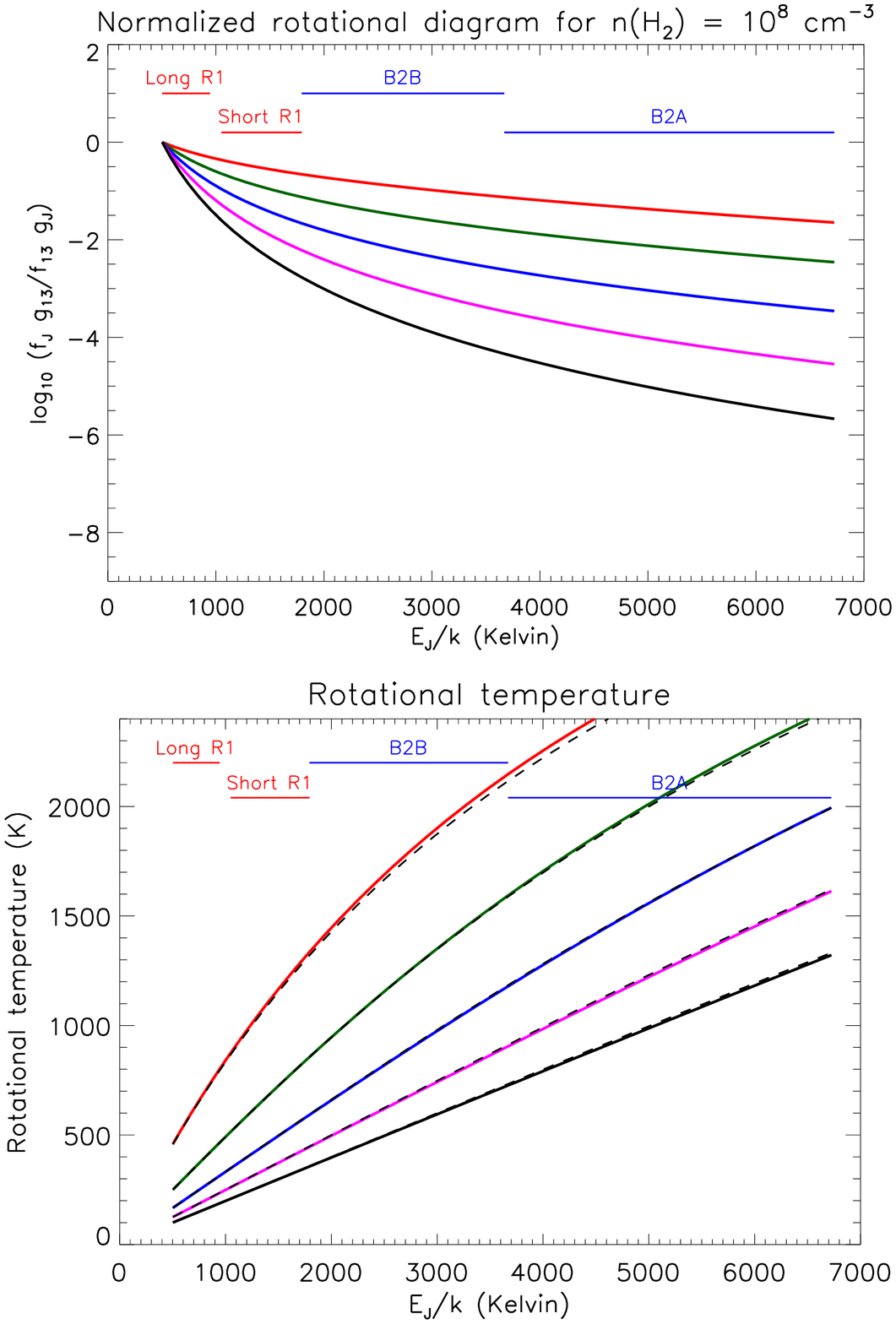}

\noindent{Fig.\ 11 -- Rotational diagrams (top panel) and rotational temperatures (bottom panel) predicted for an optically-thin medium, with a power-law distribution of gas temperatures, at an $\rm H_2$ density of $10^8\,{\rm cm}^{-3}$.  Solid black, red, green, blue and magenta curves apply to power-law indices, $b$ (see text), of 1, 2, 3, 4, and 5 respectively.  Results from an approximate analytic treatment (equation 9) are shown as dashed curves.}
\end{figure}

\vfill\eject
\begin{figure}
\includegraphics[scale=0.7,angle=0]{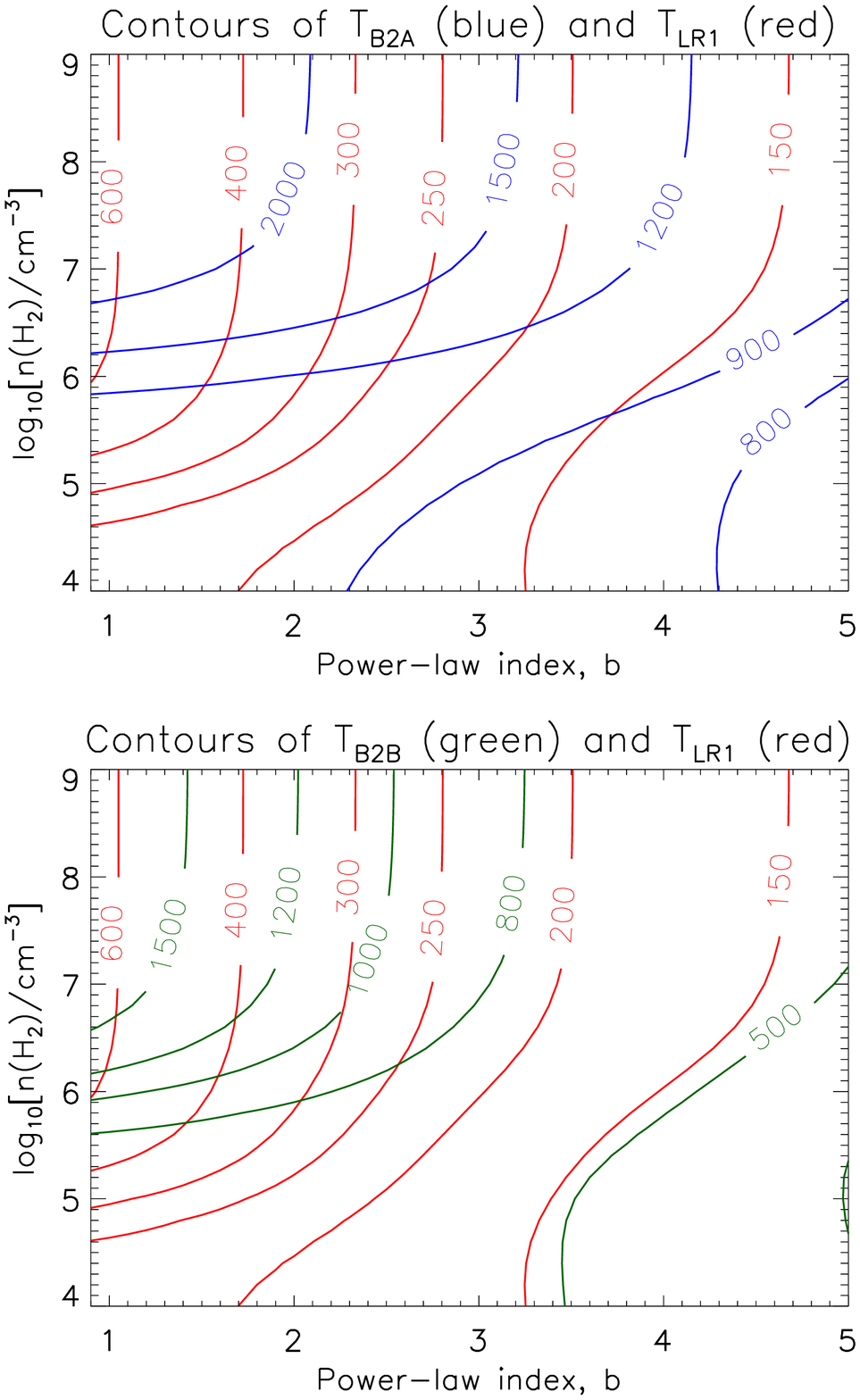}

\noindent{Fig.\ 12 --  Contour plots of the average rotational temperatures (in Kelvin) predicted for CO transitions in the Long R1 (140 -- 220~$\mu$m), B2B (70-105~$\mu$m), and B2A (51-73~$\mu$m) bands of {\it Herschel}/PACS, as a function of the power-law index, $b$, and the H$_2$ density. These are shown in red ($T_{\rm LR1}$), green ($T_{\rm B2B}$), and blue ($T_{\rm B2A}$), respectively. Results apply to an optically-thin medium with a power-law distribution of gas temperatures.}
\end{figure}

\vfill\eject
\begin{figure}
\includegraphics[scale=0.7,angle=0]{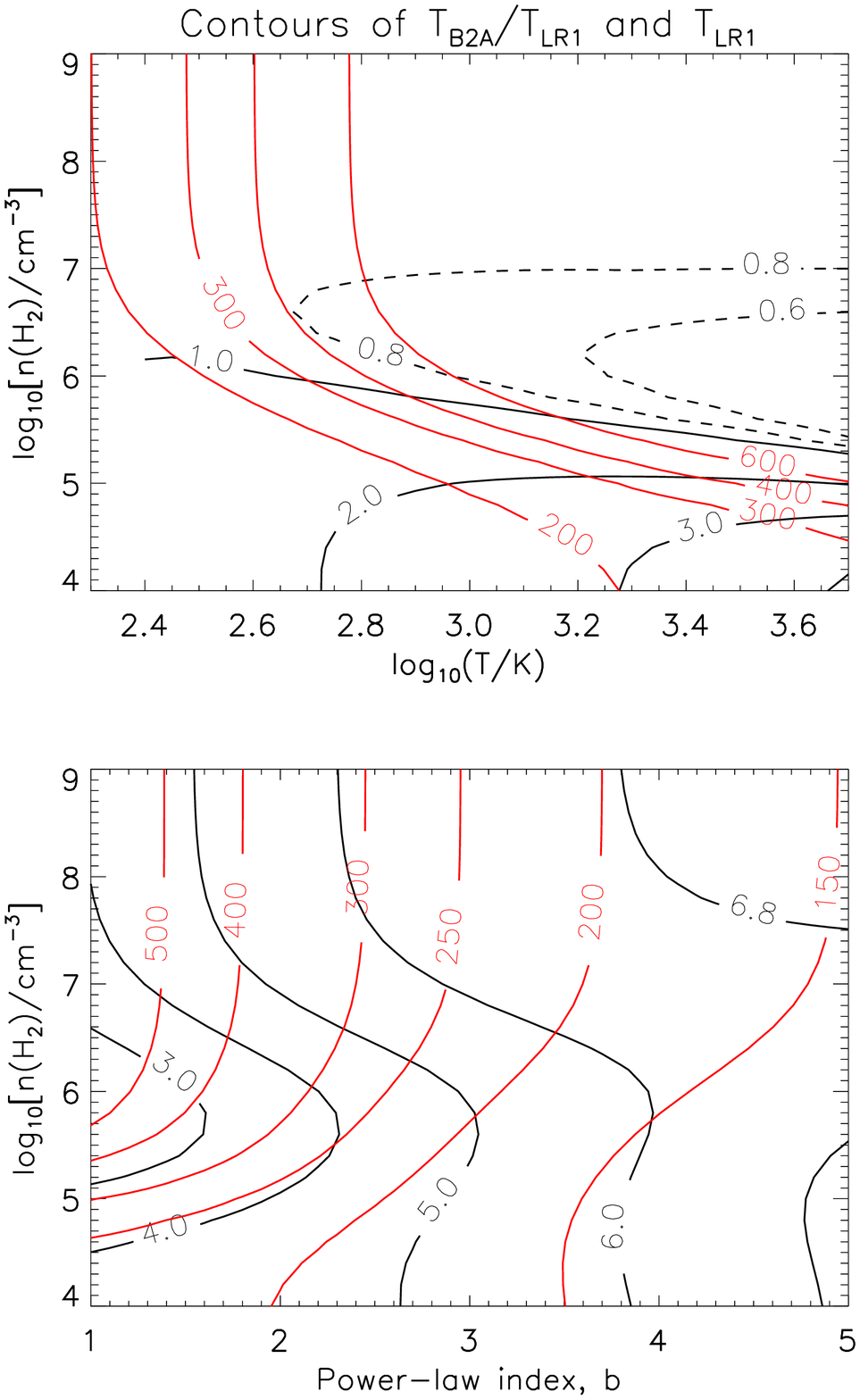}

\noindent{Fig.\ 13 -- Red contours: average rotational temperatures (in Kelvin) predicted for CO transitions in the PACS Long R1 band (red).  Black contours: ratio of average rotational temperature in the PACS B2A band to that in the Long R1 band.  Results apply to an optically-thin medium at a single temperature (top panel) or with a power-law distribution (see text) of gas temperatures (bottom panel).}
\end{figure}


\begin{thebibliography}

\bibitem[Depristo et al.(1979)]{1979JChPh..71..850D} Depristo, A.~E., 
Augustin, S.~D., Ramaswamy, R., \& Rabitz, H.\ 1979, \jcp, 71, 850

\bibitem[Fich et 
al.(2010)]{2010A&A...518L..86F} Fich, M., Johnstone, D., van Kempen, T.~A., et al.\ 2010, \aap, 518, L86 

\bibitem[Flower 
\& Gusdorf(2009)]{2009MNRAS.395..234F} Flower, D.~R., \& Gusdorf, A.\ 2009, \mnras, 395, 234 

\bibitem[Fixsen(2009)]{2009ApJ...707..916F} Fixsen, D.~J.\ 2009, \apj, 707, 
916 

\bibitem[Giannini et al.(2011)]{2011ApJ...738...80G} Giannini, T., Nisini, 
B., Neufeld, D., et al.\ 2011, \apj, 738, 80 

\bibitem[Goldflam et al.(1977)]{1977JChPh..67.5661G} Goldflam, R., Kouri, 
D.~J., \& Green, S.\ 1977, \jcp, 67, 5661 

\bibitem[Goorvitch(1994)]{1994ApJS...95..535G} Goorvitch, D.\ 1994, \apjs, 
95, 535 

\bibitem[Herczeg et al.(2011)]{2011arXiv1111.0774H} Herczeg, G.~J., Karska, 
A., Bruderer, S., et al.\ 2011, arXiv:1111.0774 

\bibitem[Jankowski 
\& Szalewicz(2005)]{2005JChPh.123j4301J} Jankowski, P., \& Szalewicz, K.\ 2005, \jcp, 123, 104301 

\bibitem[Lerate et al.(2006)]{2006MNRAS.370..597L} Lerate, M.~R., Barlow, 
M.~J., Swinyard, B.~M., et al.\ 2006, \mnras, 370, 597 

\bibitem[Mauersberger et 
al.(1988)]{1988A&A...201..123M} Mauersberger, R., Wilson, T.~L., \& Henkel, C.\ 1988, \aap, 201, 123 

\bibitem[McKee et al.(1982)]{1982ApJ...259..647M} McKee, C.~F., Storey, 
J.~W.~V., Watson, D.~M., \& Green, S.\ 1982, \apj, 259, 647 

\bibitem[Neufeld 
\& Kaufman(1993)]{1993ApJ...418..263N} Neufeld, D.~A., \& Kaufman, M.~J.\ 1993, \apj, 418, 263 

\bibitem[Neufeld 
\& Melnick(1991)]{1991ApJ...368..215N} Neufeld, D.~A., \& Melnick, G.~J.\ 1991, \apj, 368, 215 

\bibitem[Neufeld et al.(2006)]{2006ApJ...649..816N} Neufeld, D.~A., 
Melnick, G.~J., Sonnentrucker, P., et al.\ 2006, \apj, 649, 816 

\bibitem[Neufeld 
\& Yuan(2008)]{2008ApJ...678..974N} Neufeld, D.~A., \& Yuan, Y.\ 2008, \apj, 678, 974 

\bibitem[Nisini et al.(2010)]{2010ApJ...724...69N} Nisini, B., Giannini, 
T., Neufeld, D.~A., et al.\ 2010, \apj, 724, 69 

\bibitem[Poglitsch et 
al.(2010)]{2010A&A...518L...2P} Poglitsch, A., Waelkens, C., Geis, N., et al.\ 2010, \aap, 518, L2 


\bibitem[Sch{\"o}ier et 
al.(2005)]{2005A&A...432..369S} Sch{\"o}ier, F.~L., van der Tak, F.~F.~S., van Dishoeck, E.~F., \& Black, J.~H.\ 2005, \aap, 432, 369 

\bibitem[van Kempen et 
al.(2010)]{2010A&A...518L.128V} van Kempen, T.~A., Green, J.~D., Evans, N.~J., et al.\ 2010a, \aap, 518, L128 

\bibitem[van Kempen et 
al.(2010)]{2010A&A...518L.121V} van Kempen, T.~A., Kristensen, L.~E., Herczeg, G.~J., et al.\ 2010b, \aap, 518, L121 

\bibitem[Varshalovich 
\& Khersonskii(1977)]{1977ApL....18..167V} Varshalovich, D.~A., \& Khersonskii, V.~K.\ 1977, \aplett, 18, 167 

\bibitem[Viscuso 
\& Chernoff(1988)]{1988ApJ...327..364V} Viscuso, P.~J., \& Chernoff, D.~F.\ 1988, \apj, 327, 364 

\bibitem[Visser et al.(2011)]{2011arXiv1110.4667V} Visser, R., Kristensen, 
L.~E., Bruderer, S., et al.\ 2011, arXiv:1110.4667 

\bibitem[Watson et al.(1985)]{1985ApJ...298..316W} Watson, D.~M., Genzel, 
R., Townes, C.~H., \& Storey, J.~W.~V.\ 1985, \apj, 298, 316 

\bibitem[Yang et al.(2010)]{2010ApJ...718.1062Y} Yang, B., Stancil, P.~C., 
Balakrishnan, N., \& Forrey, R.~C.\ 2010, \apj, 718, 1062 

\bibitem[Yuan 
\& Neufeld(2011)]{2011ApJ...726...76Y} Yuan, Y., \& Neufeld, D.~A.\ 2011, \apj, 726, 76 


\end{thebibliography}
\end{document}